\documentclass{llncs}

\title{Integrated IoT and Cloud Environment for Fingerprint Recognition}

\usepackage{graphicx}
\usepackage[colorinlistoftodos,prependcaption,textsize=tiny]{todonotes}
\usepackage{blindtext}
\presetkeys%
{todonotes}%
{inline,backgroundcolor=yellow}{}

\usepackage{caption}
\usepackage{subcaption}
\usepackage{subcaption}
\author{Ehsan Nadjaran Toosi \inst{1}, Adel Nadjaran Toosi \inst{1}, Reza Godaz \inst{2}, \and Rajkumar Buyya \inst{1} }

\institute{
	Cloud Computing and Distributed Systems (CLOUDS) Laboratory\\
	School of Computing and Information Systems \\
	The University of Melbourne, Australia\\
	\email{\{enadjaran,anadjaran,rbuyya\}@unimelb.edu.au}\and
	Department of Software Enegineering \\
	Islamic Azad University of Mashhad\\
	\email{rgodaz@mshdiau.ac.ir}
}

\begin{document}

\maketitle

\begin{abstract}
Big data applications involving the analysis of large datasets becomes a critical part of many emerging paradigms such as smart cities, social networks and modern security systems.  Cloud computing has developed as a mainstream for hosting big data applications by its ability to provide the illusion of infinite resources.  However, harnessing cloud resources for large-scale big data computation is application specific to a large extent. In this paper, we propose a system for large-scale fingerprint matching application using Aneka, a platform for developing scalable applications on the Cloud. We present the design and implementation of our proposed system and conduct experiments to evaluate its performance using resources from Microsoft Azure.  Experimental results demonstrate that matching time for biometric information such as fingerprints in large-scale databases can be reduced substantially using our proposed system.
\end{abstract}

\section{Introduction}
Big data applications are getting popular in many fields due to the quick expansion of the Internet, smart cities, Internet of Things (IoT) devices in producing data~\cite{TOOSI2018}. In most of the cases, the data requires being processed and structured for further procedures~\cite{chen2014data}. Processing big data is often very time-consuming while it could be decreased by increasing the computation power. One of the most preferred approaches to speed up big data processing is cloud computing~\cite{buyya2009cloud}. 

Cloud computing can provide an infinite amount of computing, storage, and network resources which suits big data challenges. The data could also be stored entirely in a local infrastructure and only transferred to public infrastructure for more computation power while the trade-off between data transfer and computation power need to be considered.

In this paper, we propose a design and implementation of a big data and security-based application for searching and finding a matched fingerprint as a biometric information among a massive database of fingerprints records. The main aim of the application is to find personal information attached to the matched fingerprint. For instance, we suppose that a police department is responsible for finding the information of a person whose fingerprint has been found in a crime scene rapidly in a massive database of records. However, there are two challenges against this goal. The first is that the local computation power is limited and the number of records is enormous. The second is to compare a pair of fingerprints, the features of fingerprints are needed to be extracted and compared which is a cumbersome computational task. 

Our proposed implementation aims at utilizing the computation power of hybrid or multi-cloud. In this regard, we utilize a middleware framework, named Aneka \cite{vecchiola2012deadline}, which is a Platform-as-a-Service (PaaS) solution and provides Application Programming Interfaces (APIs) for the developers to deploy their applications. We build and deploy a finger matching application on Azure cloud resources using Aneka Software Development Kit (SDK). For fingerprint verification, we use a framework presented in~\cite{medina2014introducing}.

The rest of the paper is organized as follows. Section \ref{section:related-work} discusses the related work. Section \ref{section:architechture} represents the system architecture and how it works. Section \ref{section:system-design-and-implementation} defines the main modules and interface of the system in detail. We evaluate and analyze the performance of our system in section \ref{section:performance-and-design-evaluation}. Finally, section \ref{section:conclusion-and-future-work} is dedicated to the conclusion and future works.

\section{Related Work}\label{section:related-work}
A similar work to ours is fast fingerprint identification for large databases~\cite{peralta2014fast} where authors proposed a distributed framework for fingerprint matching for large databases. Similarly, their framework is also flexible to any kind of fingerprint matching algorithms. Le et al.~\cite{le2016complete} and Cappelli et al.~\cite{cappelli2015large} benefit from graphics processing unit (GPU) computing implementation of a fingerprint matching algorithm to speed up their system performance on large databases. These algorithms have been implemented completely using CUDA~\cite{nickolls2008scalable} where they have used different parallelization approaches. CUDA is a parallel computing framework which enables software developers to use GPU for general purpose processing. This approach is called General-Purpose computing on Graphics Processing Units (GPGPU). Even though the performance gain is high, the implementation of the parallel algorithms in CUDA is relatively hard and cumbersome. In this paper, we used Aneka platform and its easy-to-use and high-level APIs~\cite{vecchiola2012deadline} which give us the power to concentrate on task distributions and make it flexible to use different fingerprint matching algorithms.

Many studies in cloud computing have addressed the expansion of local infrastructure capacity by using public cloud resources. Toosi et al.~\cite{TOOSI2018} focus on data-intensive applications and consider the impact of data transfers in their decision for using local or public infrastructure. Mateescu et al.~\cite{mateescu2011hybrid} propose a High-Performance Computing (HPC) infrastructure architecture to execute scientific applications. Assun{\c{c}}{\~a}o et al.~\cite{de2010cost} examine the usage cost and the performance of different public cloud resource provisioning algorithms. Belgacem and Chopard~\cite{belgacem2015hybrid} conduct an empirical study of running a massively parallel MPIs application over an existing HPC infrastructure and bursting into Amazon EC2 clusters if the need arises. They aim to evaluate the overhead of using public cloud resources. Yuan et al.~\cite{yuan2017temporal} intend to utilize the temporal variation of prices public clouds to maximize profit in a hybrid cloud environment.

\section{System Architecture}\label{section:architechture}
\begin{figure}	
	\centering
	\includegraphics[width=1.0\textwidth]{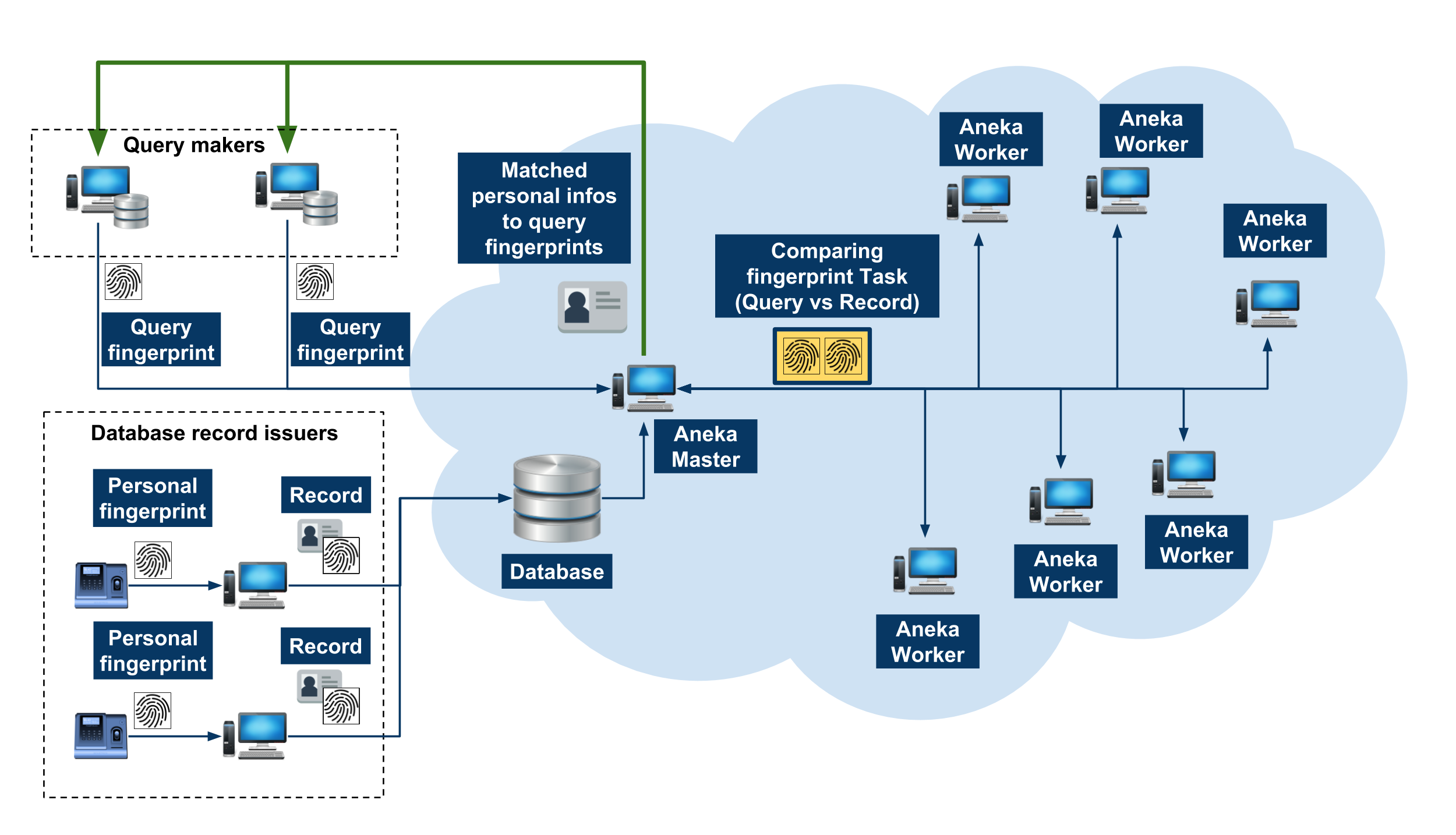}
	\caption{System architecture.}
	\label{fig:system-architechture}
\end{figure}
This section provides a general overview of how the entire system works. Figure \ref{fig:system-architechture} visualizes the system architecture. The system contains three components which are database record issuers, query makers, and cloud-based fingerprint searching.

The \textit{database record issuers} is the part of the system which provide the input data for the system. The peoples' fingerprints are binded with their personal information (i.e personal photo and name and etc.) and stored in the database. For the simplicity of the system, we use a file-system to store all records.

The second part of the system is the \textit{query makers}. Queries are requested to find the matched fingerprints to the query fingerprint of interest. In this regard, requests are sent to the main component of the system which is \textit{cloud-based fingerprint searching}.

The cloud-based fingerprint searching is controlled by Aneka cloud platform. Considering that the number of records in the database and the number of queries can be huge, the importance of parallel searching in the database is obvious. The requests are given to the Aneka master (main node) which is responsible to make and distribute the comparison tasks among Aneka workers. Each comparison task consists all query fingerprints and a single fingerprint record in the database. The total number of comparison task is equal to the number of database records and independent to the number of queries. Aneka workers return matched similarity index between the fingerprint database record and query fingerprints to Aneka master. Aneka master gathers all the similarity indexes computed by Aneka workers, finds the maximum similarity, retrieves the personal information of matched fingerprints to query fingerprints and returns them to the query makers.

Two main components of the used framework for fingerprint recognition which are computationally heavy are 1) extracting features from fingerprint images and 2) comparing fingerprint features. To avoid re-extracting feature of both query and database fingerprints in the searching procedure, Aneka master extracts the feature of query fingerprints, and Aneka workers extract the features of database fingerprint records and compare them against the query fingerprint features.

\section{System Design and Implementation}\label{section:system-design-and-implementation}
\begin{figure}[t!]
	\centering
	\includegraphics[width=0.85\textwidth]{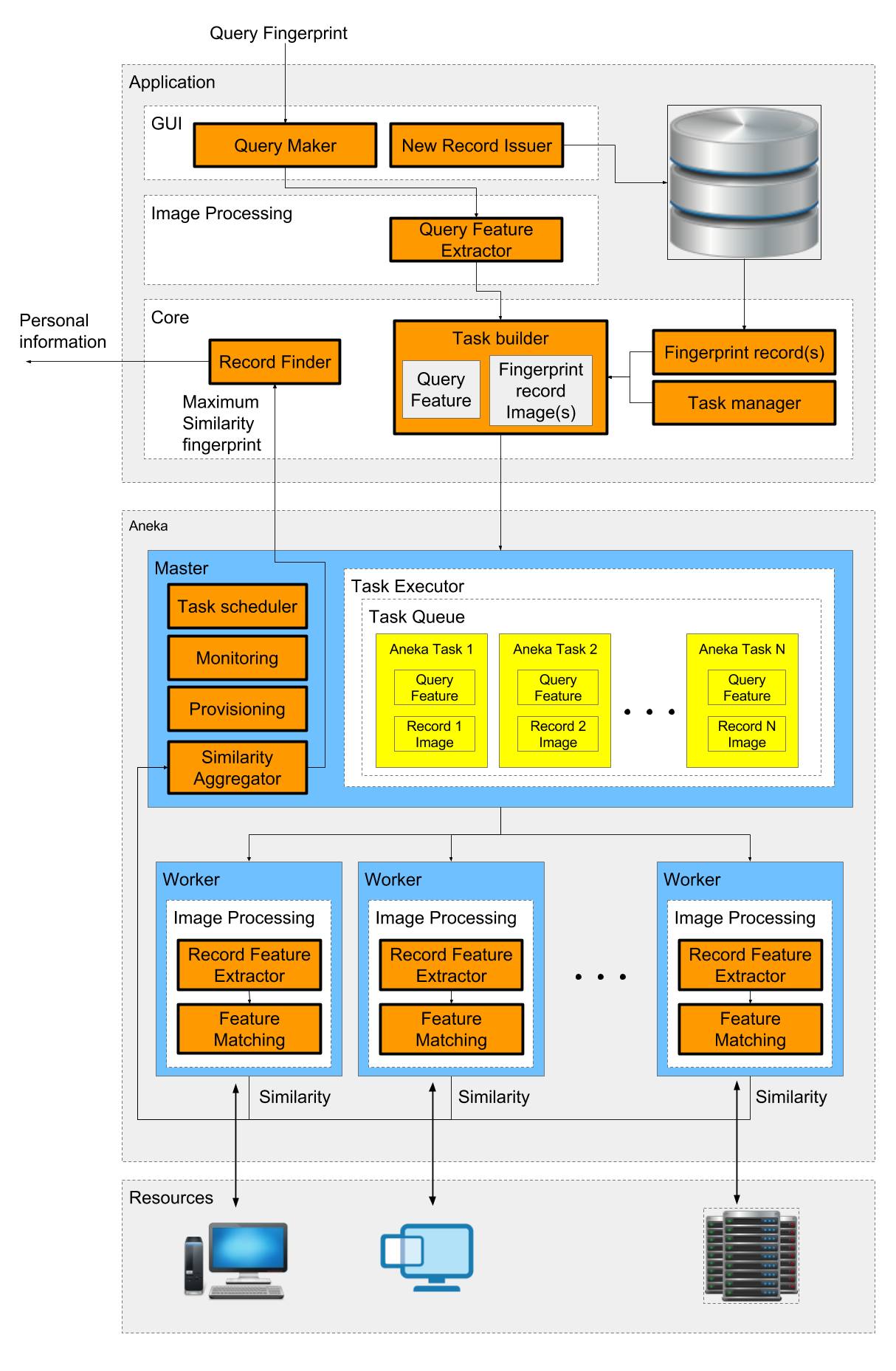}
	\caption{System design and implementation.}
	\label{fig:system-design-and-implementation}
\end{figure}

Figure \ref{fig:system-design-and-implementation} shows a layered view of our system's key components. It also provides the data flow in the system. The layered system design contains three layers.

The top layer belongs to the fingerprint recognition application and its main functionalities. The user is able to make a new record to store in the database or search a fingerprint through the database to retrieve the information of the person who is matched with the query fingerprint. Upon the receipt of a request for fingerprint matching, the features of the query fingerprint are extracted. Task Manager is the part of the system which tells how many database fingerprint records are needed to be packed in a comparison task. For the sake of simplicity, in this paper, we only pack one record in each comparison task. Task builder is the component that makes comparison tasks. It uses Aneka ITask interface to prepare the comparison task for the master. Later on, master submits Tasks to the workers for the execution.


The middle layer belongs to the Aneka cloud which performs as a middleware providing computational resources to the fingerprint matching application. Aneka tasks which are created previously in the application layer are buffered in the task queue in the Aneka master. Aneka master submits Aneka tasks to the Aneka workers for comparing record and query fingerprints. After receiving Aneka tasks by Aneka workers, they start to extract the features of record in the task, and compare it to the query fingerprint features. Finally, they return the matching similarity index to the master node to aggregate results.

The bottom layer provides computational resources for the Aneka platform to execute its tasks. Resources residing in this layer are obtained and managed by Aneka from a variety of sources, including private and public clouds, clusters, grids, and desktop grids.

\begin{figure}[t!]
	\centering
	\includegraphics[width=0.6\textwidth]{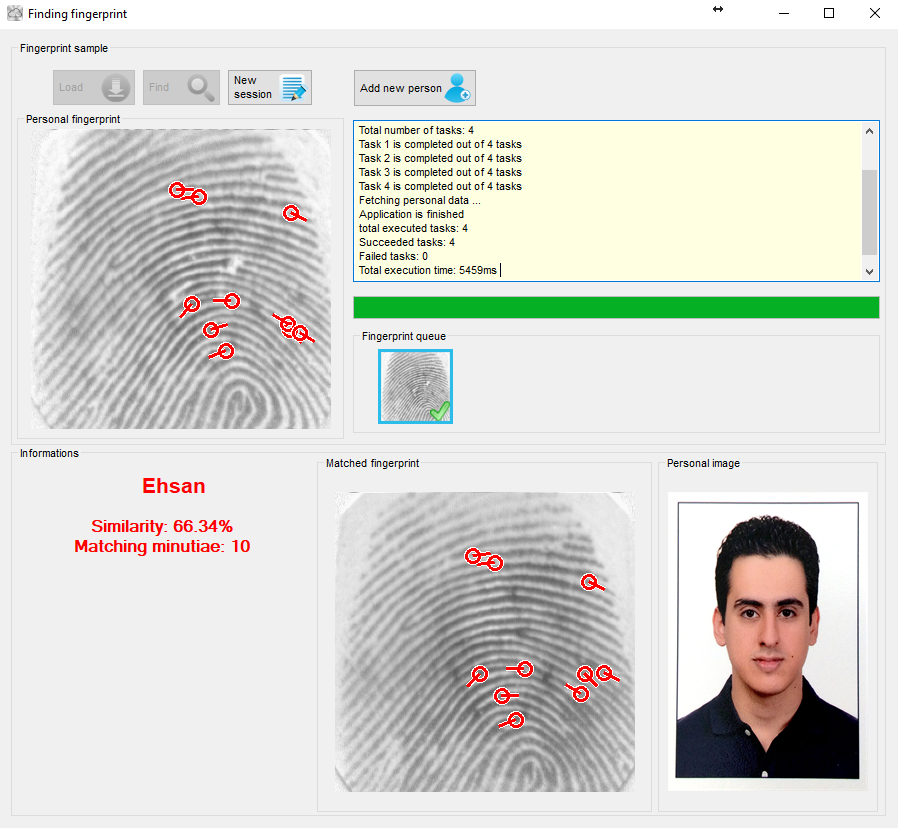}
	\caption{Application GUI}
	\label{fig:application-gui}
\end{figure}

\section{Performance Evaluation}\label{section:performance-and-design-evaluation}
\begin{figure}[t!]
	\centering
	\begin{minipage}{.45\textwidth}
		\includegraphics[width=1.0\textwidth]{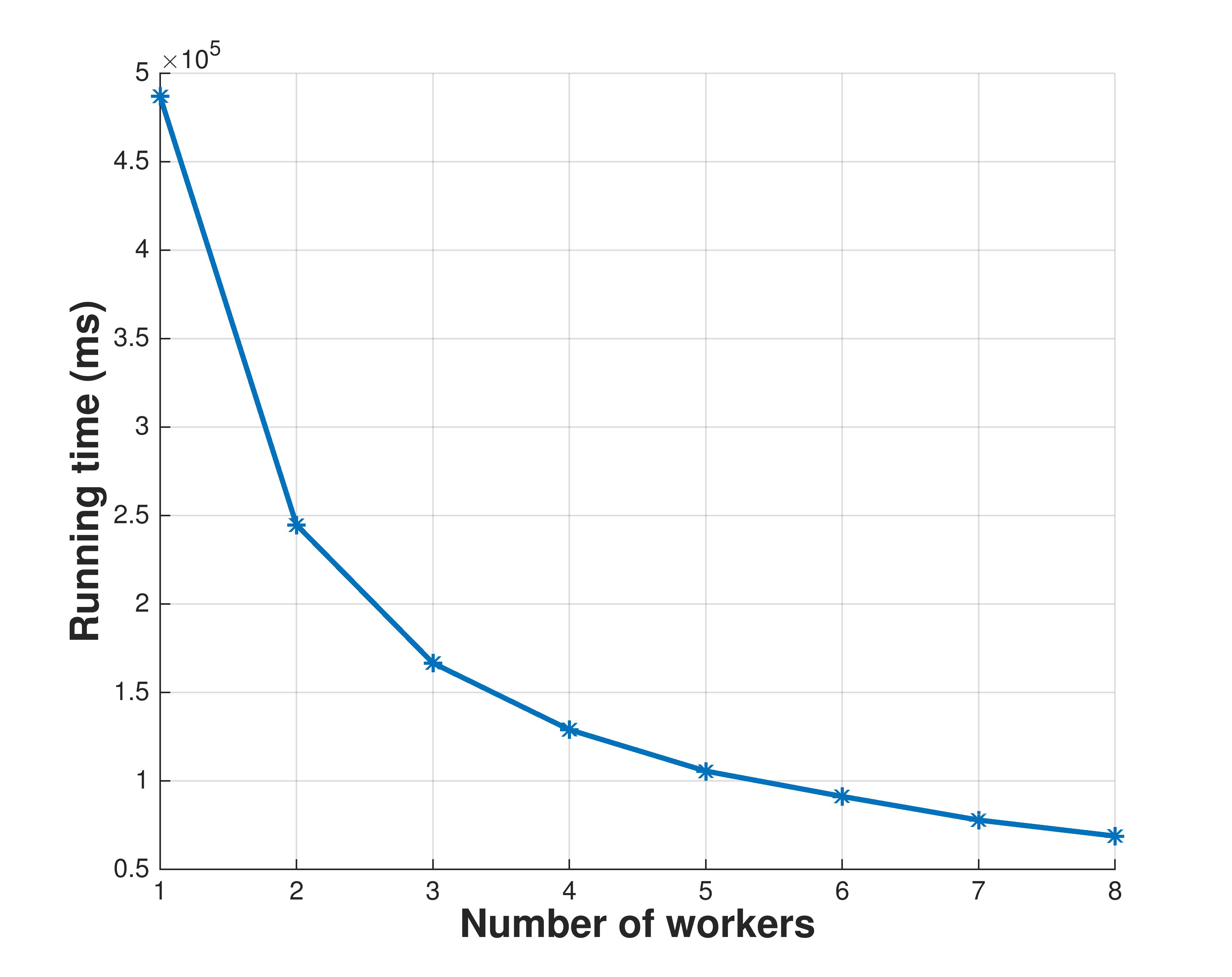}
		\caption{Running time vs number of workers}
		\label{fig:running-time-vs-number-of-workers}
	\end{minipage}
	\begin{minipage}{.45\textwidth}
		\includegraphics[width=1.0\textwidth]{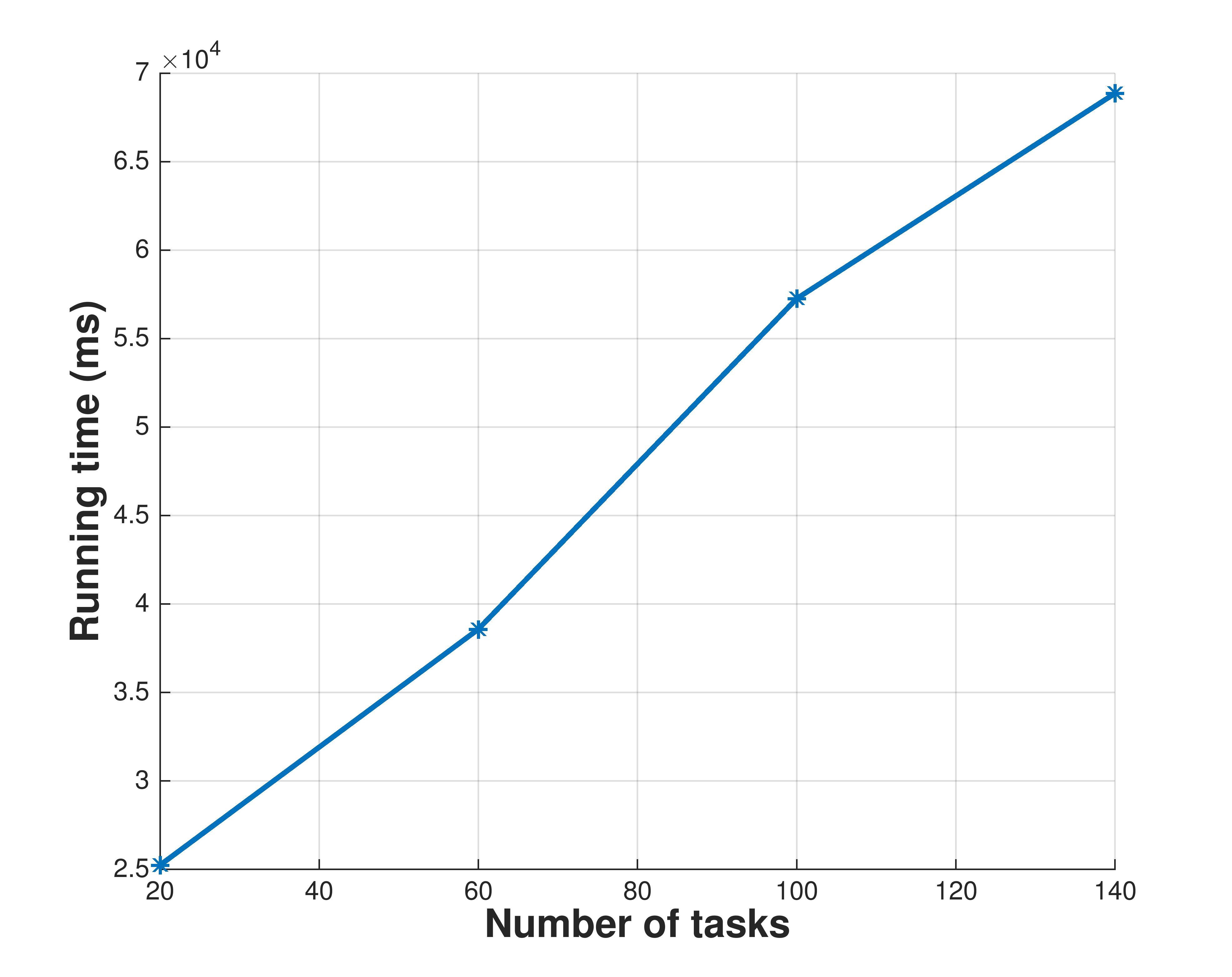}
		\caption{Running time vs number of tasks}
		\label{fig:running-time-vs-number-of-tasks}
	\end{minipage}
\end{figure}
In this section, we evaluate the performance of the system in terms of running time using two different experiments. For this purpose, we used the GUI of our application shown in Figure~\ref{fig:application-gui}. Query fingerprints are queued for finding the matched person. The personal information of the matched person along with the matched fingerprint are displayed. There is also a console showing the progress of fingerprint searching such as tasks distribution, total running time, etc.

Both experiments are run on a master and a set of worker machines. The master runs on a desktop machine residing at the University of Melbourne and workers are provisioned from the Microsoft Azure Australia Southeast region. The master machine is an Intel Core i5-430M (2 Cores and 4 Logical Processors) at 2.27GHz, 8GB main memory and runs under Microsoft Windows 10 Pro operating system. Worker machines are single core Azure Instances (Standard DS1) with a 2.4GHz processor and 3.5GB main memory running Windows Server 2012 as the operating system. We configured our application task manager to create an Aneka task (comparing task) per each fingerprint in the database.

In the first experiment, the number of tasks is fixed to 140 and on the other hand, the running time of the application is evaluated by varying the number of workers from 1 to 8. As expected, increasing the number of workers reduces the running time of the system. For instance, Figure \ref{fig:running-time-vs-number-of-workers} shows that the running time of the system for a single and two workers are almost 500 and 250 seconds, respectively, where the running time is reduced to half. Eventually, the running time reaches about 69 seconds when the number of workers is increased to 8.

In the second experiment, the number of workers is fixed to 8 and the running time is analyzed by changing the number of working tasks to 20, 60, 100 and 140. Figure \ref{fig:running-time-vs-number-of-tasks} displays that the corresponding running time is nearly 25, 40, 57 and 68 seconds which is demonstrating a linear growth in time versus the number of tasks.



\section{Conclusion and Future Work}\label{section:conclusion-and-future-work}
In this paper, we demonstrated the benefits of using cloud computing for fingerprint recognition and matching for a large-scale database. We presented the design and implementation of our system including its architecture. We showed that how Aneka provides the required platform for scheduling and parallel execution of tasks on public cloud resources, e.g., Azure. A conducted performance evaluation showed that the fingerprint queries could be responded in a significantly lower timeframe using our proposed system.

As a future work, we are planning to devise a technique for dynamic resource provisioning based on the number of queries. The future direction will be to extend our system as a Software-as-a-Service (SaaS), one of the major categories of cloud computing, for the security-oriented organizations.

\bibliographystyle{IEEEtran}
\bibliography{references}

\end{document}